\documentclass[aps,prc,reprint]{revtex4-2}
\usepackage{graphicx}
\usepackage[caption=false]{subfig}

\usepackage{amssymb,amsbsy,amsfonts,amsmath}
\usepackage{epsfig,color}
\usepackage{epstopdf}
\usepackage[citecolor=blue,colorlinks=true,linkcolor=blue,urlcolor=blue]{hyperref}
\usepackage{bm}
\usepackage{braket}
\usepackage{amsmath}

\begin{document}

%\preprint{}

\title{Nonlinear polarization effects on plasma screening for fusion reactions}
\author{Hanxiang Huang$^{1}$}

\author{Binbing Wu$^{2}$}
\email{wubb17@mail.bnu.edu.cn}

\author{Zhengfeng Fan$^{2,4}$}
\email{fan_zhengfeng@iapcm.ac.cn}

\author{Congzhang Gao$^{2}$}

\author{Jie Liu$^{3,4}$}

\author{Baisong Xie$^{1,5}$}
\email{bsxie@bnu.edu.cn}

\affiliation{
$^{1}$Key Laboratory of Beam Technology of the Ministry of Education, and School of Physics and Astronomy, Beijing Normal University, Beijing 100875, People's Republic of China
\\$^{2}$Institute of Applied Physics and Computational Mathematics, Beijing 100088,People's Republic of China
\\$^{3}$Graduate School, China Academy of Engineering Physics, Beijing 100193, People's Republic of China
\\$^{4}$CAPT, HEDPS, and IFSA Collaborative Innovation Center of MoE, Peking University, Beijing 100871, People's Republic of China
\\$^{5}$Institute of Radiation Technology, Beijing Academy of Science and Technology, Beijing 100875, People's Republic of China
}


% \author{}
% %\email[]{Your e-mail address}
% %\homepage[]{Your web page}
% %\thanks{}
% %\altaffiliation{}
% \affiliation{}


\date{\today}
\begin{abstract}
Using a finite-temperature two-center Thomas-Fermi-Dirac model, we investigate nonlinear plasma screening effects on the fusion reactions D-T, $^{12}$C-$^{12}$C, and p-$^{11}$B. The effective interaction between a reacting ion pair is obtained self-consistently at the mean-field level. Combining this potential with a complex Woods-Saxon nuclear potential, we solve the stationary Schr\"odinger equation to evaluate tunneling probabilities and reaction rates. Compared with Debye-H\"uckel theory, the two-center screening potential can be stronger or weaker, and the resulting fusion enhancement correspondingly amplified or suppressed, depending on the plasma conditions and the internuclear separation. We identify the underlying mechanisms: relative to the single-center prediction, the nonlinear evacuation of background ions suppresses screening, whereas nonlinear electron accumulation enhances it. The two channels respond to different plasma state parameters: the ionic suppression deepens with increasing coupling strength, whereas the electronic enhancement peaks at intermediate degeneracy. The two-center correction therefore follows a factorized scaling law in the nuclear charges of the reacting pair and the plasma state parameters. These findings highlight the importance of a self-consistent two-center treatment for determining effective interactions and evaluating static screening corrections to fusion reaction rates.
\end{abstract}



\maketitle

\section{Introduction}

In plasmas, the long-range Coulomb repulsion between reacting ions is screened by surrounding charges, thereby enhancing the quantum tunneling probability and, consequently, the fusion reaction rates \cite{salpeter1954electrons,assenbaum1987effects}. Therefore, plasma screening is important in both stellar nucleosynthesis and controlled fusion schemes \cite{clayton1983principles, rolfs1988cauldrons,adelberger2011solar}, such as inertial confinement fusion (ICF) \cite{lindl1995development,betti2016inertial,abu2024achievement}. These extreme environments span from moderately to strongly coupled regimes (i.e., Coulomb-to-kinetic energy ratio $\Gamma \gtrsim 1$) and from partial to extreme electron degeneracy (i.e., thermal-to-Fermi energy ratio $\theta \lesssim 1$) \cite{gasques2005nuclear}. For instance, ICF implosions operate in the moderately coupled and partially degenerate regime ($\Gamma \sim 1$, $\theta \lesssim 1$) \cite{hu2011first,zylstra2022burning}, while stellar carbon burning lies in the even more strongly coupled and strongly degenerate regime ($\Gamma \gg 1$, $\theta \ll 1$) \cite{ichimaru1993nuclear}. Due to the interplay of many-body and quantum degeneracy effects, accurately modeling the plasma screening under such conditions remains a challenge \cite{grayson2024self}.

Several screening models have been used for different fusion reactions. In the classical, weakly coupled and non-degenerate limit, the Debye-H{\"u}ckel (DH) theory solves the linearized Poisson-Boltzmann (PB) equation for a single ion \cite{debye1923theory,salpeter1954electrons}. To extend the description to the degenerate regime, the Stewart-Pyatt (SP) model incorporates a finite-temperature Thomas-Fermi model \cite{stewart1966lowering}. By naturally accounting for electron degeneracy, the SP model evaluates the continuum lowering, successfully bridging the gap between the Debye limit and the ion-sphere limit. Subsequently, the Mitler model evaluates the Helmholtz free energy difference within a simplified Thomas-Fermi model, thereby yielding a unified screening potential and corresponding enhancement factors across a wide density range \cite{mitler1977thermonuclear,gruzinov1998screening}. For a fully quantum-mechanical treatment of the screening cloud, the average-atom model self-consistently solves the Kohn-Sham equations for a single atom embedded in a plasma \cite{rozsnyai1972relativistic,liberman1979self,starrett2013electronic}. These approaches rest on a single-center assumption: they describe the screening cloud around individual ions, neglecting the mutual polarization when two nuclei approach each other. As a result, they cannot accurately evaluate the effective two-body interaction in the many-body plasma. In principle, first-principles simulations such as Path-Integral Monte Carlo \cite{militzer2001path, pollock2004dense} and Quantum Molecular Dynamics \cite{collins1995quantum} can self-consistently capture many-body correlations and quantum degeneracy. However, their computational cost is prohibitive, and their statistical noise is severe at the short internuclear distances relevant for quantum tunneling. Alternatively, in strongly coupled and fully degenerate plasmas, the two-center Ion-Sphere (IS) model has been widely employed to account for the strongly coupled screening of the reacting pair, by assuming that two isolated electron drops merge into a single united drop \cite{salpeter1954electrons,salpeter1969nuclear,kravchuk2014strong}.

Building upon the physical picture of the two-center IS model \cite{salpeter1954electrons}, we develop a two-center Thomas-Fermi-Dirac (TC-TFD) screening model that generalizes the framework of Mitler \cite{mitler1977thermonuclear}. At the self-consistent mean-field level, the model captures the nonlinear two-center polarization of the plasma, providing a more effective description of two-body interactions in the many-body system than single-center treatments. By replacing the simplified assumption of uniform electron drops with self-consistent spatial distributions, our model extends the two-center framework to a finite-temperature TFD description \cite{eichler1974two} whose ion and electron treatments are designed to span the $\Gamma-\theta$ plane. Incorporating the Fermi-Dirac distribution and the Dirac exchange potential \cite{dirac1930note,feynman1949equations}, our model yields the self-consistent shared screening cloud around two colliding nuclei. The effective two-body interaction potential, i.e., the variation of the system's Helmholtz free energy with internuclear separation, is obtained by integrating the mean force acting on the nuclei along the internuclear axis \cite{kirkwood1935statistical}. Combining this potential with a complex Woods-Saxon nuclear potential, we solve the stationary Schr\"odinger equation by the Numerov method \cite{johnson1978renormalized} to obtain tunneling probabilities and reaction rates for deuterium–tritium (D-T), carbon 12–carbon 12 ($^{12}$C-$^{12}$C), and proton–boron 11 (p-$^{11}$B) reactions. 

Our results reveal that the TC-TFD screening potential can be either stronger or weaker than the DH prediction, and the resulting fusion enhancement correspondingly amplified or suppressed, depending on the plasma conditions and the internuclear separation. We identify this behavior as the competition between the nonlinear polarization of ions and that of electrons: relative to the single-center prediction, the saturation of background-ion evacuation suppresses the ionic screening contribution, whereas the nonlinear accumulation of electrons enhances it. The two channels respond to distinct plasma state parameters: the ionic suppression deepens with increasing coupling strength $\Gamma$, whereas the electronic enhancement reaches a pronounced peak near the quantum--classical boundary $\theta \approx 1$. Furthermore, the two-center correction obeys a factorized scaling law in the nuclear charges of the reacting pair and the plasma state parameters. Because the Gamow tunneling window shifts to shorter internuclear distances as the temperature rises, the interplay between far-field ionic suppression and near-field electronic enhancement leads to a non-monotonic temperature dependence in the fusion reaction-rate enhancement. These findings highlight the importance of a self-consistent two-center treatment for determining effective interaction potentials and evaluating static screening corrections to thermonuclear fusion rates.

The paper is organized as follows. Section II presents the theoretical framework of the TC-TFD model and the numerical methods for the self-consistent field and the fusion reaction rate. Section III compares the TC-TFD screening potential with the single-center TFD (SC-TFD), DH, PB, and IS models, decomposes it into its ionic and electronic channels, systematically demonstrates the scaling laws with respect to nuclear charges and plasma parameters ($\Gamma$ and $\theta$), and evaluates the corresponding fusion cross-section and reaction-rate enhancement factors. Section IV concludes with implications for dense plasmas relevant to inertial confinement fusion and gives an outlook.

\section{Theoretical framework}
\label{sec:theory}
\subsection{The TC-TFD screening model}

\begin{figure}[t]
    \centering
    \includegraphics[width=0.90\linewidth]{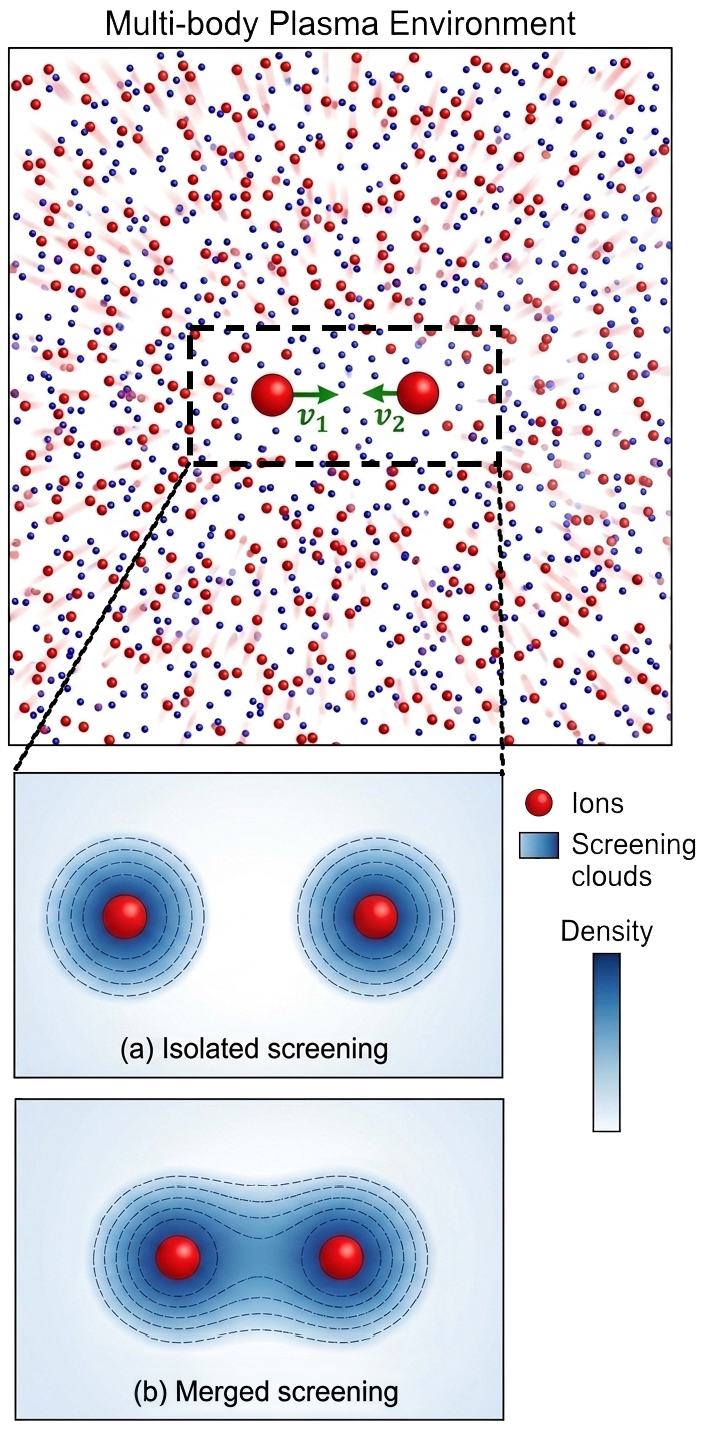}
    \caption{Schematic illustration of two nuclei approaching each other in a plasma with velocities \(\vec{v}_1\) and \(\vec{v}_2\) in a multi-body plasma environment. The top panel shows the microscopic background composed of background ions (red spheres) and electrons (blue dots), with the black dashed box indicating the central interaction region. The panel $(\text{a})$ shows their isolated, spherically symmetric screening clouds before contact; the panel $(\text{b})$ illustrates the deformation and merging of screening clouds as the nuclei collide at short internuclear distances. Dashed lines indicate charge density contours.}
    \label{fig.1}
\end{figure}

A plasma is a complex many-body system, in which the ensemble average of many-body interactions manifests as a screening cloud that effectively reduces the bare Coulomb potential around nuclei. During the fusion reaction process, as the colliding nuclei approach each other within the Debye length, the individual screening clouds interact, deform, and eventually merge into a shared screening cloud under the influence of their combined electric field \cite{kravchuk2014strong, furutani1993two}, as illustrated in Fig.~\ref{fig.1}. The equilibrium structure of this shared screening cloud is nontrivial due to the combined Coulomb potential of the two approaching nuclei and the nonlinear response of the background plasma. Therefore, a statistical thermodynamic approach is required to evaluate the effective interaction of the colliding nuclei \cite{kirkwood1935statistical,salpeter1969nuclear,ichimaru1993nuclear}.

Consider an $N$-body plasma system where the two colliding nuclei are located at positions $\mathbf{r}_1$ and $\mathbf{r}_2$, and the surrounding plasma consists of $N-2$ background particles. The classical configurational partition function of the system is given by:
\begin{align}
    \mathcal{Z} = \int \exp\left[ -\frac{U(\mathbf{r}_1, \dots, \mathbf{r}_N)}{k_\text{B} T} \right] d\mathbf{r}_1 \dots d\mathbf{r}_N,
    \label{eq.1}
\end{align}
where $U(\mathbf{r}_1, \dots, \mathbf{r}_N)$ is the total Coulomb potential energy among all particles. This potential energy can be partitioned into pairwise interactions:
\begin{align}
    U(\mathbf{r}_1, \dots, \mathbf{r}_N) = &V_{\text{c}}(|\mathbf{r}_1 - \mathbf{r}_2|) + \sum_{i=1}^{2}\sum_{j=3}^{N} V_\text{c}(\mathbf{r}_i,\mathbf{r}_j) \notag \\
    &+ \sum_{3 \le k < l \le N} V_\text{c}(\mathbf{r}_k,\mathbf{r}_l),
    \label{potential}
\end{align}
where $V_\text{c}(\mathbf{r}_i,\mathbf{r}_j) = \frac{Z_i Z_j e^2}{4 \pi \epsilon_0 |\mathbf{r}_i - \mathbf{r}_j|}$ is the Coulomb potential between particles $i$ and $j$, and $Z_i$ is the charge number of particle $i$. 

To evaluate the effective interaction between the two colliding nuclei, one can trace out the environmental degrees of freedom by integrating over the coordinates of the $N-2$ background particles, yielding the two-particle reduced partition function $\mathcal{Z}_2(\mathbf{r}_1,\mathbf{r}_2)$:
\begin{align}
    \mathcal{Z}_2(\mathbf{r}_1, \mathbf{r}_2) = \int \exp\left[ -\frac{U(\mathbf{r}_1, \dots, \mathbf{r}_N)}{k_\text{B} T} \right] d\mathbf{r}_3 \dots d\mathbf{r}_N.
    \label{eq.2}
\end{align}
Without loss of generality, by fixing nucleus 1 at the coordinate origin ($\mathbf{r}_1 = \mathbf{0}$) and nucleus 2 at $\mathbf{r}_2 = \mathbf{r}$, the translational and rotational invariance of the isotropic background plasma ensures that $\mathcal{Z}_2$ depends strictly on the scalar internuclear distance $r = |\mathbf{r}_2 - \mathbf{r}_1|$, which we denote simply as $\mathcal{Z}_2(r)$.

In statistical mechanics, this reduced partition function is related to the pair distribution function $g(r)$ of the colliding nuclei via $g(r) = \frac{V^2}{\mathcal{Z}} \mathcal{Z}_2(r)$, where $V$ is the system volume. According to Kirkwood's formalism, this spatial distribution defines the potential of mean force (PMF) via the relation $V_{\text{pmf}}(r) = -k_\text{B} T \ln g(r)$ \cite{kirkwood1935statistical,dewitt1973screening,graboske1973screening}, which represents the effective two-body interaction potential under the ensemble average of the many-body environment \cite{kirkwood1935statistical,henderson1974uniqueness,hansen2013theory}. Furthermore, the logarithm of the reduced partition function defines the reduced Helmholtz free energy of the system, $F_{\text{reduced}}(r) = -k_\text{B} T \ln \mathcal{Z}_2(r)$. At infinite separation ($r \to \infty$), the spatial correlation between the two nuclei vanishes, giving the asymptotic decorrelation condition $\lim_{r \to \infty} g(r) = 1$. This condition establishes the zero reference of the effective interaction, rendering the PMF equivalent to the variation of the Helmholtz free energy as the nuclei approach each other from infinity:
\begin{align}
V_{\text{pmf}}(r) = F_{\text{reduced}}(r) - F_{\text{reduced}}(\infty).
\label{eq:pmf_work}
\end{align}
This free energy variation corresponds to the reversible work $W^{\text{rev}}_{\infty \rightarrow r}$ required to bring the two nuclei together in the plasma.

Building upon this, the screening potential $V_{\text{screen}}(r)$ is obtained by subtracting the PMF from the bare Coulomb potential \cite{salpeter1954electrons, ichimaru1993nuclear}, which equals the reduction of the system's free energy due to the polarization of the background \cite{onsager1933theories,onsager1939electrostatic}:
\begin{align}
V_{\text{screen}}(r) &= V_{\text{c}}(r) - V_{\text{pmf}}(r) \notag \\
&= -\Delta F_{\text{bg}}(r),
\label{eq:screening_potential}
\end{align}
where $\Delta F_{\text{bg}}(r)$ is the change in the Helmholtz free energy of the background plasma.

Direct calculation of the subtle change in total Helmholtz free energy is highly susceptible to numerical round-off errors. To circumvent this difficulty, we instead evaluate the reversible work by integrating the mean force—the negative spatial gradient of the PMF—exerted on the reacting nuclei along the internuclear axis. Taking the gradient with respect to the relative position vector $\mathbf{r}$ of nucleus 2 (with charge $Z_2 e$) yields:
\begin{align}
    -\nabla_{\mathbf{r}} V_{\text{pmf}}(r) &= k_\text{B} T \frac{\nabla_{\mathbf{r}} \mathcal{Z}_2(r)}{\mathcal{Z}_2(r)} \notag \\
    &= \frac{k_\text{B} T}{\mathcal{Z}_2(r)} \int \nabla_{\mathbf{r}} \exp\left[-\frac{U(\mathbf{0},\mathbf{r},\mathbf{r}_3,\dots,\mathbf{r}_N)}{k_\text{B} T}\right] d\mathbf{r}_3 \dots d\mathbf{r}_N \notag \\
    &= \int (-\nabla_{\mathbf{r}} U) \frac{\exp\left[ -\frac{U}{k_\text{B} T} \right]}{\mathcal{Z}_2(r)} d\mathbf{r}_3 \dots d\mathbf{r}_N \notag \\
    &\equiv \langle -\nabla_{\mathbf{r}} U \rangle_{N-2}.
    \label{eq:mean_force_derivation}
\end{align}
Using Eq.~\eqref{potential}, the mean force takes the explicit form:
\begin{align}
    \langle -\nabla_{\mathbf{r}} U \rangle_{N-2} &= -Z_2 e \nabla_{\mathbf{r}} \left\langle \frac{Z_1 e}{4\pi\epsilon_0 r} + \sum_{j=3}^{N} \frac{Z_j e}{4\pi\epsilon_0 |\mathbf{r} - \mathbf{r}_j|} \right\rangle_{N-2} \notag \\
    &= -Z_2 e \nabla_{\mathbf{r}} \Phi(\mathbf{r}),
\end{align}
where $\Phi(\mathbf{r})$ represents the total electrostatic potential field acting on nucleus 2, generated by nucleus 1 and the ensemble-averaged charge distribution of the background plasma \cite{feynman1939forces}. Due to the axial symmetry along the internuclear path, this ensemble-averaged force acts strictly along the internuclear axis $\hat{\mathbf{r}}$, defining the scalar mean force $\mathcal{F}(r)$:
\begin{align}
    \mathcal{F}(r) = \langle -\nabla_{\mathbf{r}} U \rangle_{N-2} \cdot \hat{\mathbf{r}} = -Z_2 e \frac{\partial \Phi(\mathbf{r})}{\partial r}.
    \label{eq:macro_force}
\end{align}
The potential of mean force $V_{\text{pmf}}(r)$ is then obtained by integrating the mean force $\mathcal{F}(r')$ along the internuclear axis from infinity to $r$:
\begin{align}
V_{\text{pmf}}(r) = - \int_{\infty}^{r} \mathcal{F}(r') \, \text{d} r'.
\label{eq:pmf_integral}
\end{align}
We emphasize that Eqs.~\eqref{eq.1}--\eqref{eq:pmf_integral} constitute the classical statistical-mechanical definition of the PMF. In evaluating it, we do not perform the full $N$-body trace; instead, following Mermin's density-functional theory \cite{teller1962stability,mermin1965thermal}, we map the background plasma response onto a finite-temperature TFD grand-potential functional \cite{feynman1949equations,cowan1957extension,eichler1974two}. The explicit expressions for the grand potential functional are detailed in Appendix \ref{sec:tfd_model}. The resulting $V_{\rm pmf}$ therefore captures the two-center nonlinear polarization of the background at the self-consistent mean-field level.

The question now reduces to evaluating the self-consistent electrostatic potential $\Phi(\mathbf{r})$. Minimizing the TFD grand-potential functional determines the local electron density $n_e(\mathbf{r})$ and ion density $n_i(\mathbf{r})$; coupling these charge distributions with the Poisson equation yields the following set of nonlinear self-consistent field equations. For given nuclear positions at $\mathbf{r}_1$ with charge $Z_1e$ and $\mathbf{r}_2$ with charge $Z_2e$, we obtain:
\begin{align}
&\nabla^{2}\Phi(\mathbf{r}) = -\frac{e}{\epsilon_0} \Bigl( \sum_{j=1,2} Z_{j}\,\delta^3(\mathbf{r}-\mathbf{r}_j) \notag \\
&\qquad + \sum_{i} z_{i} n_{i}(\mathbf{r}) - n_{e}(\mathbf{r}) \Bigr),
\label{eq:poisson}\\
&n_{e}(\mathbf{r}) = n_{e}(\infty) 
\frac{F_{1/2}\!\left( \frac{\mu_{e}+e\Phi(\mathbf{r})-V_{\text{x}}(\mathbf{r})}{k_\text{B} T} \right)}
{F_{1/2}\!\left( \frac{\mu_{e}-V_{\text{x}}(\infty)}{k_\text{B} T} \right)},
\label{eq:ne}\\
&n_{i}(\mathbf{r}) = n_{i}(\infty) \exp\!\left[-\frac{z_{i}e\Phi(\mathbf{r})}{k_\text{B} T}\right],
\label{eq:ni}\\
&V_{\text{x}}(\mathbf{r}) = -\frac{e^2}{4\pi\epsilon_0} \left( \frac{3 n_e(\mathbf{r})}{\pi} \right)^{1/3}.
\label{eq:vex}
\end{align}
In these equations, the delta-function terms in Eq.~\eqref{eq:poisson} account for the two-center nature of the problem, and $\Phi(\mathbf{r})$ is the total electrostatic potential generated by both the nuclei and the polarized plasma. Equation~\eqref{eq:ne} expresses the local electron density via the Fermi-Dirac integral $F_{1/2} $\cite{mohankumar1995accurate,blakemore1982approximations} and the local Dirac exchange potential $V_{\text{x}}$ \cite{dirac1930note}, with the electron chemical potential $\mu_{e}$ determined asymptotically by $n_e(\infty)=A_e F_{1/2}\!\left( (\mu_{e}-V_\text{x}(\infty))/k_\text{B} T \right)$, where $A_e = \sqrt{2}(m_e k_\text{B} T)^{3/2}/(\pi^2 \hbar^3)$ is the density-of-states prefactor. Concurrently, each ion species $i$ follows its classical Boltzmann distribution. A single temperature, $T_e=T_i=T$, is assumed throughout. 

We iteratively solve the nonlinear self-consistent field equations using a finite-difference method on a two-dimensional cylindrical grid to exploit azimuthal symmetry. Because the relevant spatial scales range from the macroscopic Debye length at the boundaries down to the subatomic scale of the internuclear separation, a non-uniform grid is essential. We employ a recursive, two-way coupled grid-nesting algorithm to solve the self-consistent field equations \cite{berger1984adaptive, berger1989local, brandt1977multi}. The method is described in Appendix~\ref{app:numerical_method}. This approach captures the nonlinear variations of the potential near the nuclei while maintaining computational efficiency at larger macroscopic distances. 

The TC-TFD model can be reduced to several well-known models under asymptotic limits. First, by neglecting the mutual polarization of the colliding nuclei and considering only the single isolated nucleus 1 with higher charge $Z_1 e$, the system reduces to a single-center model. This naturally yields the single-center TFD (SC-TFD) model, where the Poisson equation simplifies to a single-center source:
\begin{align}
\nabla^2\Phi_1(\mathbf{r}) = -\frac{e}{\epsilon_0}\bigg( Z_1 \delta^3(\mathbf{r}) + \sum_{i} z_i n_i(\mathbf{r}) - n_e(\mathbf{r}) \bigg).
\label{eq:sctfd}
\end{align}
Here, the densities $n_e(\mathbf{r})$ and $n_i(\mathbf{r})$ retain their exact finite-temperature forms as defined in Eqs.~\eqref{eq:ne} and \eqref{eq:ni}. To construct the screening potential in this single-center system, nucleus 1 is fixed at the origin while the approaching nucleus 2 is treated as a bare point test charge $Z_2 e$ at distance $r$ that samples the unperturbed field of the central ion. Decomposing the single-center field into the bare nuclear potential and the plasma-induced potential, $\Phi_1(r) = \frac{Z_1 e}{4\pi\epsilon_0 r} + \Phi_1^{\text{ind}}(r)$, the single-center PMF and screening potential are defined as:
\begin{align}
V_{\text{pmf}}^{\text{SC}}(r) = Z_2 e\, \Phi_1(r), \quad V_{\text{screen}}^{\text{SC}}(r) = -Z_2 e\, \Phi_1^{\text{ind}}(r).
\label{eq:screen_sc}
\end{align}
Furthermore, in the non-degenerate limit, the quantum exchange potential becomes negligible ($V_{\text{x}} \to 0$), and the Fermi integral asymptotically recovers the classical Boltzmann distribution. Under this classical limit, the SC-TFD model reduces to the nonlinear Poisson-Boltzmann (PB) model:
\begin{align}
\nabla^2\Phi(\mathbf{r}) =& -\frac{e}{\epsilon_0}\bigg( Z_1 \delta^3(\mathbf{r}) + \notag \\
&\sum_{i} z_i n_i(\infty) e^{-\frac{z_i e\Phi(\mathbf{r})}{k_\text{B} T}} - n_e(\infty) e^{\frac{e\Phi(\mathbf{r})}{k_\text{B} T}} \bigg).
\label{eq:pb}
\end{align}
Finally, in the weakly coupled regime where the electrostatic potential energy is negligible compared to the thermal energy ($|e\Phi| \ll k_\text{B} T$), linearizing the exponential terms in Eq.~\eqref{eq:pb}  directly recovers the standard Debye-Hückel (DH) model:
\begin{align}
\nabla^2\Phi(\mathbf{r}) - \kappa_\text{D}^2 \Phi(\mathbf{r}) = -\frac{Z_1 e}{\epsilon_0} \delta^3(\mathbf{r}),
\label{eq:dh}
\end{align}
where $\kappa_\text{D} = \left[ \frac{e^2}{\epsilon_0 k_\text{B} T} \left( \sum_i z_i^2 n_i(\infty) + n_e(\infty) \right) \right]^{1/2}$ is the inverse Debye length.

\subsection{Tunneling probabilities and fusion reaction rates}

Using the screening potential $V_{\text{screen}}(r)$ obtained from the previous section, we calculate the tunneling probabilities and fusion reaction rates. In the center-of-mass (CM) frame, the fusion process can be described by a stationary Schr\"odinger equation in the relative coordinate system \cite{wu2025multiphoton, wu2024multiphoton, thomson2024laser}. The reduced radial wave function is given by $\psi(r) = rR(r)$, where $R(r)$ is the radial part of the original three-dimensional wave function, and the radial variable $r$ corresponds to the internuclear distance \cite{atzeni2004physics}. For low energy reactions, we only consider the $s$-wave scattering, which satisfies \cite{clayton1983principles,atzeni2004physics}:
\begin{align}
-\frac{\hbar^2}{2m_\text{r}}\psi''(r) + V_{\text{total}}(r)\psi(r) = E\psi(r),
\label{eq:schrodinger}
\end{align}
where $E$ is the CM kinetic energy, and $m_\text{r}$ is the reduced mass of the reacting nuclei. The total potential $V_{\text{total}}(r)$ is composed of the complex Woods-Saxon nuclear potential and the effective Coulomb potential $V_{\text{total}}(r) = V_{\text{WS}}(r) + V_{\text{pmf}}(r)$. The effective Coulomb potential is the PMF defined in Eq.~\eqref{eq:pmf_work}, which reduces to the bare Coulomb potential in the absence of plasma screening. The Woods-Saxon potential is given by:
\begin{align}
&V_{\text{WS}}(r) = -\frac{V_0 + iW_0}
{1 + \exp\!\bigl(\frac{r-R_0}{a}\bigr)},
\label{eq:woods_saxon}
\end{align}
where $V_0$ and $W_0$ are the depths of the real nuclear potential and the imaginary absorption potential, respectively, $R_0= r_0 (A_1^\frac{1}{3}+A_2^\frac{1}{3})$ is the nuclear radius, where $r_0=1.2 \text{fm}$, and $a=0.5 \text{fm}$ is the surface diffuseness parameter. The imaginary part $W_{\text{abs}}(r) = W_0 / \bigl(1 + e^{(r-R_0)/a}\bigr)$ accounts for irreversible nuclear absorption once the Coulomb barrier is penetrated \cite{etde_5540661, hagino2012subbarrier,gasques2005nuclear}. We adopt a representative real potential depth of $V_0=30 \text{ MeV}$ and a short-range imaginary absorption depth of $W_0=500 \text{keV}$, ensuring complete flux absorption in the interior nuclear region without causing unphysical reflections \cite{hagino2012subbarrier}. We have verified that the enhancement factors, being ratios of tunneling probabilities, are insensitive to these parameters.

By employing the Numerov method to numerically solve the stationary Schr\"odinger equation with either the bare Coulomb potential or the screened Coulomb potential, the penetration probability is obtained as the ratio of absorbed probability flux $J_{\text{abs}}$ to incident probability flux $J_{\text{in}}$ \cite{gasques2005nuclear}:
\begin{align}
P(E) = \frac{J_{\text{abs}}}{J_{\text{in}}} = \frac{\frac{2}{\hbar} \int_{0}^{R_0} W_{\text{abs}}(r)\,|\psi(r)|^2 \, \text{d}r}{\frac{\hbar k}{m_\text{r}} |C_{\text{in}}|^{2}},
\label{eq:penetration}
\end{align}
where $k = \sqrt{2m_\text{r} E}/\hbar$ is the wave number, and $C_{\text{in}}$ is the incident wave amplitude determined by matching the numerical wave function to the analytic asymptotic stationary wave functions at large distances. 

Based on the Gamow picture, the energy-dependent fusion cross-section $\sigma(E)$ is factorized into the astrophysical $S$-factor $S(E)$, the kinematic geometrical factor $1/E$, and the Coulomb barrier penetration probability $P(E)$ \cite{gamow1928quantentheorie,meyer1998inertial, rolfs1988cauldrons}:
\begin{equation}
\sigma(E) = \frac{S(E)}{E} P(E),
\label{eq:cross_section}
\end{equation} 
Because the plasma environment only alters the long-range electrostatic interactions, the screening effect leaves both the nuclear $S$-factor and the geometrical cross-section unchanged \cite{assenbaum1987effects}. Consequently, the impact of plasma screening on the total fusion cross-section can be exclusively determined by evaluating the modification of the quantum tunneling probability $P(E)$. Therefore, the enhancement factor for cross-section is defined as:
\begin{align}
f_{\mathrm{cs}}(E) = \frac{P_{\mathrm{screened}}(E)}{P_{\mathrm{bare}}(E)},
\label{eq:tunnel_enhance}
\end{align}
where both $P_{\mathrm{screened}}$ and $P_{\mathrm{bare}}$ are calculated using the Numerov method \cite{johnson1978renormalized}.

\begin{figure*}[t]
    \centering
    \includegraphics[width=1\textwidth]{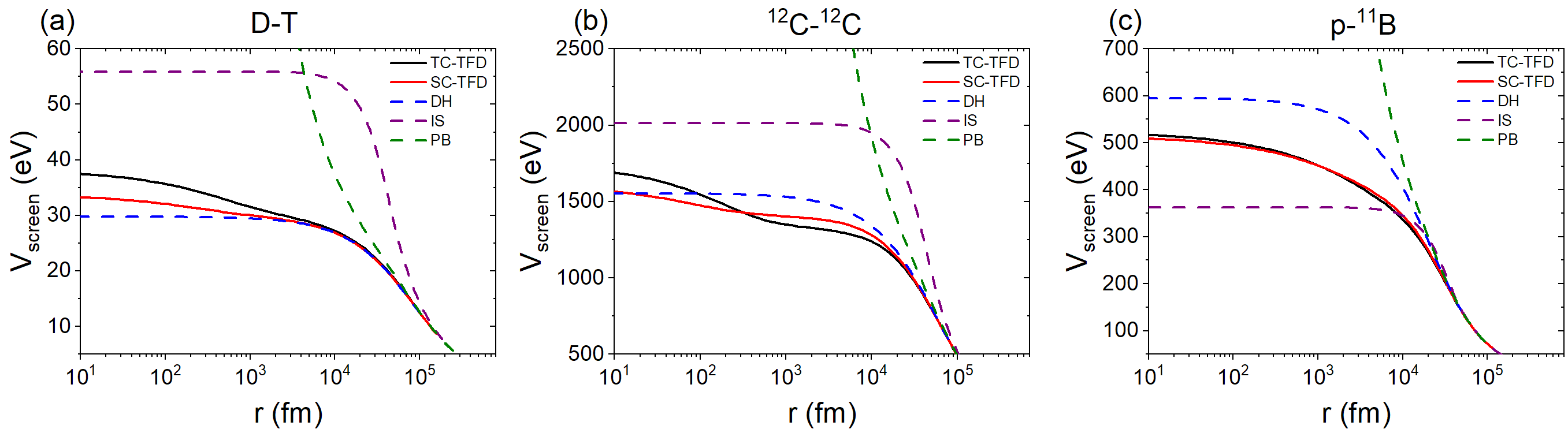}
    \caption{ (Color online) Comparison of the screening potentials $V_\text{screen}=V_\text{bare}-V_\text{pmf}$ as a function of the internuclear distance $r$ for three plasma conditions. In each panel, solid lines denote the TC-TFD (black) and the SC-TFD (red). Dashed lines represent classical and analytical results: the weak-coupling Debye-H\"{u}ckel model (blue), the classically divergent nonlinear Poisson-Boltzmann model (green), and the strong-coupling merged Ion-Sphere model (purple). Specific parameters are: (a) D-T plasma at $T = 1\,\mathrm{keV}$ with mass densities $\rho_{\text{D}} = 20\,\mathrm{g/cm^3}$ and $\rho_{\text{T}} = 30\,\mathrm{g/cm^3}$;(b) $^{12}$C-$^{12}$C plasma at $T = 10\,\mathrm{keV}$ with a total mass density of $\rho_{\text{C}} = 240\,\mathrm{g/cm^3}$; (c) p-$^{11}$B plasma at $T = 1\,\mathrm{keV}$ with $\rho_{\text{p}} = 20\,\mathrm{g/cm^3}$ and $\rho_{\text{B}} = 220\,\mathrm{g/cm^3}$.}
    \label{Different_models}
\end{figure*}

Finally, the fusion reaction rate $\langle \sigma v \rangle$ in a thermal plasma is computed by averaging over the Maxwell-Boltzmann distribution \cite{atzeni2004physics}:
\begin{align}
\langle \sigma v \rangle
=& \frac{4}{(2\pi m_\text{r})^{1/2}} \frac{1}{(k_\text{B} T)^{3/2}}\notag \\
&\times\int_0^{\infty} \sigma(E) \, E \, e^{-E/(k_\text{B} T)} \, \text{d}E,
\label{eq:rate}
\end{align}
the enhancement factor for the fusion reaction rate is defined as
\begin{align}
f_{\text{rate}} = \frac{\langle \sigma v \rangle_{\text{screened}}}{\langle \sigma v \rangle_{\text{bare}}},
\label{eq:rate_enhance}
\end{align}
which provides a macroscopic measure of the plasma screening effect. We emphasize that the thermal reaction rate $\langle \sigma v \rangle _\text{screened}$ relies on the static TC-TFD screening potential evaluated earlier. We follow the static-screening approach of Assenbaum, Langanke and Rolfs \cite{assenbaum1987effects}, inserting the equilibrium potential of mean force as an $r$-dependent effective potential in the radial Schr\"odinger equation. Its validity rests on the assumption that the background remains in local equilibrium throughout the nuclear approach. In full thermal equilibrium, the reaction rate is determined by the equilibrium pair correlation function, so that dynamical corrections to the screening do not contribute at leading order \cite{gruzinov1998screening,brown1997nuclear}; genuinely time-dependent corrections to the ionic response lie beyond the scope of the present static treatment.

In fusion reactions, the integrand $\sigma(E)E e^{-E/k_\text{B} T}$ is sharply peaked around the Gamow peak energy. Since the astrophysical $S$-factor $S(E)$ varies slowly within this narrow window of the Gamow peak energy \cite{salpeter1954electrons, rolfs1988cauldrons}, treating $S(E)$ as a slowly varying constant that cancels out in the ratio allows us to evaluate the thermal enhancement of the fusion rate and its relative deviation from classical models.

\section{Numerical Results and Discussion} 

\subsection{Two-center screening potential and channel decomposition}

To systematically evaluate the screening effects of the TC-TFD model, we calculate the screening potential $V_{\text{screen}}(r) = V_{\text{c}}(r) - V_{\text{pmf}}(r)$ for three fusion plasmas: D-T, $^{12}$C-$^{12}$C, and p-$^{11}$B, spanning weakly to moderately coupled, non-degenerate to partially degenerate regimes, as well as symmetric and asymmetric charge configurations (Table~\ref{tab:params}). We compare the results with those of the SC-TFD and several conventional models in Fig.~\ref{Different_models}.

\begin{table}[b]
    \centering
    \begin{ruledtabular}
    \begin{tabular}{lccccc}
        Reaction & $T$ (keV) & $\rho_1$ ($\mathrm{g/cm^3}$) & $\rho_2$ ($\mathrm{g/cm^3}$) & $\Gamma$ & $\theta$  \\
        \hline
        D-T               & 1  & 20  & 30         & 0.053 & 5.29  \\
        $^{12}$C-$^{12}$C & 10 & 240 & \text{---} & 0.19  & 16.00 \\
        p-$^{11}$B        & 1  & 20  & 220        & 0.87  & 1.60  \\
    \end{tabular}
    \end{ruledtabular}
    \caption{Plasma parameters of the three benchmark reactions. $T$ is the plasma temperature, $\rho_1$ and $\rho_2$ denote the fuel mass densities, $\Gamma=\langle Z^2 \rangle e^2/(4\pi \epsilon_\text{0} a_i k_\mathrm{B}T)$ is the ion coupling parameter with $\langle Z^2\rangle=\sum_i n_i z_i^2/\sum_i n_i$ and $a_i=(3/4\pi\sum_i n_i)^{1/3}$. $\theta=k_\mathrm{B}T/E_\mathrm{F}$ is the electron degeneracy parameter, with $E_\mathrm{F}$ the Fermi energy.}
    \label{tab:params}
\end{table}

\begin{figure*}[t]
\centering
\includegraphics[width=1\linewidth]{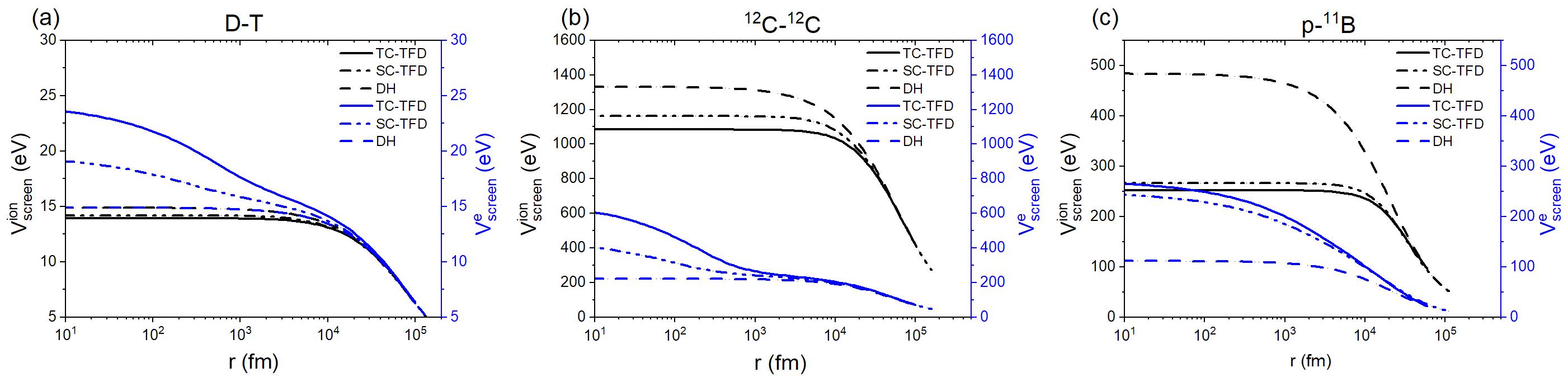}
\caption{Decomposition of the total screening potential into the contributions of electrons and ions. Black curves (left axis) denote the ionic screening potential $V_{\text{screen}}^{\text{ion}}$. Blue curves (right axis) denote the electronic screening potential $V_{\text{screen}}^{\text{e}}$. Solid, dash-dotted, and dash lines correspond to the TC-TFD, SC-TFD, and DH models, respectively. Specific parameters are: (a) D-T plasma at $T = 1\,\mathrm{keV}$ with $\rho_{\text{D}} = 20\,\mathrm{g/cm^3}$ and $\rho_{\text{T}} = 30\,\mathrm{g/cm^3}$ ; (b) $^{12}$C-$^{12}$C plasma at $T = 10\,\mathrm{keV}$ with a total mass density of $\rho_{\text{C}} = 240\,\mathrm{g/cm^3}$; (c) p-$^{11}$B plasma at $T = 1\,\mathrm{keV}$ with $\rho_{\text{p}} = 20\,\mathrm{g/cm^3}$ and $\rho_{\text{B}} = 220\,\mathrm{g/cm^3}$.}
\label{split}
\end{figure*}

As shown in Fig.~\ref{Different_models}, in the far-field region ($r \gtrsim 10^5\,\mathrm{fm}$), the weak electrostatic field ensures that all models asymptotically converge to the linear DH potential. At shorter internuclear distances, the classical PB model severely over-accumulates electrons due to the absence of Fermi statistics \cite{feynman1949equations}, while the IS model establishes a plateau representing the Salpeter strong-coupling limit \cite{salpeter1954electrons, salpeter1969nuclear}.

Focusing on the TC-TFD model and the SC-TFD model, the potential profiles reveal distinct spatial behaviors across the three reactions. In the symmetric D-T plasma [Fig.~\ref{Different_models}(a)], the TC-TFD potential exceeds the DH and SC-TFD curves at short distances due to electron accumulation, while SC-TFD remains closely aligned with DH. For $^{12}$C-$^{12}$C [Fig.~\ref{Different_models}(b)], a pronounced spatial crossover occurs: the TC-TFD potential is suppressed below DH and SC-TFD at intermediate separations ($r \gtrsim 500\,\mathrm{fm}$), but surpasses them in the near-field region. In the asymmetric p-$^{11}$B plasma [Fig.~\ref{Different_models}(c)], the TC-TFD and SC-TFD potentials remain nearly indistinguishable at all internuclear distances, with a residual of at most a few eV, and both stay below the DH prediction. 

To elucidate the mechanisms underlying these phenomena, Fig.~\ref{split} decomposes the screening potential into separate ionic and electronic contributions: $V_{\text{screen}}(r) = V_{\text{screen}}^{\text{ion}}(r) + V_{\text{screen}}^{\text{e}}(r)$, where $V_{\text{screen}}^{\text{ion}}(r)$ is the screening potential generated from the background ion polarization, and $V_{\text{screen}}^{\text{e}}(r)$ is the screening potential generated from the background electron polarization. By decomposing TC-TFD, SC-TFD, and DH model predictions, we reveal two opposing physical mechanisms acting at distinct spatial scales.

The far-field screening potential is governed by the nonlinear polarization of ions: based on Eq.~\eqref{eq:ni}, the local ion density is determined by the Boltzmann factor $e^{-z_i e\Phi(\mathbf{r}) / k_\text{B} T}$, and the exponent scales with the plasma coupling strength,
\begin{align}
\frac{z_i e\Phi(\mathbf{r})}{k_\text{B} T} = \Gamma \cdot \left[ \frac{4\pi\epsilon_0 a_i z_i \Phi(\mathbf{r})}{\langle Z^2 \rangle e} \right] \propto z_i \Gamma  \Phi(\mathbf{r}).
\label{eq:boltzmann_gamma}
\end{align}
The strong Coulomb repulsion from the approaching nuclei evacuates surrounding ions, creating a Coulomb hole around them. For ions with $z_i \ge 1$, their nonlinear polarization is more pronounced at relatively large distances, thereby dominating the far-field screening potential. More importantly, relative to the single-center results, the two-center nonlinear polarization of the background ions lowers the ionic screening potential, which can be understood from the thermodynamic free energy: the two nuclei mutually expel the background ions in the overlap area, and because the local ion density cannot drop below zero, the mutual evacuation becomes depleted toward saturation, restricting further decrease of the free energy and hence limiting the screening potential. Therefore, the TC-TFD yields a smaller screening potential than that of the SC-TFD. The magnitude of this reduction is controlled by the coupling strength $\Gamma$. As the ion coupling strength increases, the enhanced electrostatic repulsion forms a bigger Coulomb hole and expands their mutual overlap. Consequently, for a given reacting pair, the gap between the SC-TFD (black dash-dot line) and the TC-TFD models (black solid line) becomes larger in the stronger coupled plasmas. 

In contrast, the near-field screening potential is governed by the nonlinear polarization of electrons. As the two nuclei approach each other, their superimposed Coulomb potentials create a highly attractive potential well. Unlike the background ions, which are physically bounded by a minimum density of zero, the electrons respond to this strong attraction by nonlinear density accumulation. Consequently, the $V^e_\text{screen}$ for the TC-TFD model (blue solid line) is consistently higher than that of the SC-TFD (blue dash-dot line) in Fig.~\ref{split}. The magnitude of this nonlinear enhancement is governed by the background degeneracy $\theta$. In Eq.~\eqref{eq:ne}, the local density of electrons is determined by $\eta_e$, which is mainly governed by the background electron chemical potential $\mu_e$, the local self-consistent electrostatic potential $\Phi$, and the temperature $T$:
\begin{align}
    \eta_e(\mathbf{r}) = \frac{\mu_e + e\Phi(\mathbf{r})-V_\text{x}}{k_\text{B} T},
\end{align}
where $\mu_e$ is related to the $\theta$ via $F_{1/2}\left( \frac{\mu_e - V_{\text{x}}(\infty)}{k_\text{B} T} \right) = \frac{2}{3}\theta^{-3/2}$, and $\Phi(\mathbf{r})$ is primarily determined by the charges and separation of the colliding nuclei. In regions where $\eta_e(\mathbf{r}) \ll 0$, the local electrons behave classically, and nonlinear polarization enhances the accumulation without Pauli suppression. Conversely, as electrons are drawn into the strongly attractive region, the increasing $e\Phi(\mathbf{r})$ drives $\eta_e(\mathbf{r}) \gg 0$, and the system undergoes a transition where Pauli blocking is activated, limiting the degree of nonlinear accumulation. Consequently, as the electron degeneracy parameter $\theta$ decreases, the competition between electron accumulation and quantum Pauli suppression leads to a turnaround, where the electronic gap reaches a peak at intermediate degeneracy.

In symmetric collisions (D-T and $^{12}$C-$^{12}$C), the two nuclei exert comparable electrostatic forces, driving significant mutual deformation of the shared screening cloud and producing a prominent two-center gap. In contrast, for the highly asymmetric p-$^{11}$B pair, the single proton acts predominantly as a weak perturbation on the dominant screening cloud established by the central boron ion. Consequently, the single-center model captures the dominant part of the polarization for p-$^{11}$B. 

In summary, the effective screening interaction is governed by a spatial competition between far-field ionic evacuation suppression and near-field electronic accumulation enhancement. This behavior is modulated jointly by the plasma coupling strength, electron degeneracy, and nuclear charge symmetry. The quantitative scaling of these mechanisms, and their factorization into a nuclear-charge factor and a plasma-state response, are established in Sec.~III\,B.

\subsection{Scaling law of the two-center correction}

To identify the physical mechanisms governing the two-center correction and quantify the trends observed in Figs.~\ref{Different_models} and \ref{split}, we examine the nonlinear response of the background plasma within a perturbative framework. Expanding the local induced charge density $\rho_{\text{ind}}[\Phi]$ in powers of the total electrostatic potential $\Phi(\mathbf{r})$ yields:
\begin{equation}
\label{eq:rho_expand}
\rho_{\text{ind}}(\mathbf{r}) = \chi_1 \Phi(\mathbf{r}) + \chi_2 \Phi^2(\mathbf{r}) + \mathcal{O}(\Phi^3),
\end{equation}
where $\chi_1$ represents the linear screening susceptibility, and $\chi_2 = \chi_2(\Gamma, \theta)$ is the leading nonlinear susceptibility governed by the plasma coupling strength $\Gamma$ and electron degeneracy $\theta$. In the linear limit, the screened potential generated by each colliding nucleus is $\Phi_j(\mathbf{x}) = Z_j e G(\mathbf{x}, \mathbf{r}_j)$ ($j=1,2$), where $G$ denotes the screened Green's function satisfying $(\nabla^2 - \kappa_\text{D}^2)G = -\delta^3 / \epsilon_0$.

The screening potential corresponds to the change in the background free energy: $V_{\text{screen}}(r) = -\Delta F_{\text{bg}}(r)$ [Eq.~\eqref{eq:screening_potential}]. Integrating the quadratic charge density $\chi_2 \Phi^2$ against the electrostatic potential $\Phi$ yields the leading nonlinear correction to the free energy, which enters at cubic order $\sim -\frac{1}{3}\chi_2 \int \Phi^3 \text{d}^3\mathbf{x}$. Substituting the superimposed linear fields $\Phi_1 + \Phi_2$ and subtracting the distance-independent self-polarization energies ($\Phi_1^3$ and $\Phi_2^3$), the leading two-center cross-polarization contribution to the screening potential is:
\begin{align}
\label{eq:F_cross}
V_{\text{screen}}^{\text{TC}(2)}(r) &= -\chi_2 \int \left[ \Phi_1^2(\mathbf{x})\Phi_2(\mathbf{x}) + \Phi_1(\mathbf{x})\Phi_2^2(\mathbf{x}) \right] \text{d}^3\mathbf{x} \notag \\
&\propto -\chi_2 Z_1 Z_2 (Z_1 + Z_2),
\end{align}
which is symmetric under nuclear exchange ($Z_1 \leftrightarrow Z_2$).

\begin{figure}[t]
    \centering
    \includegraphics[width=0.9\columnwidth]{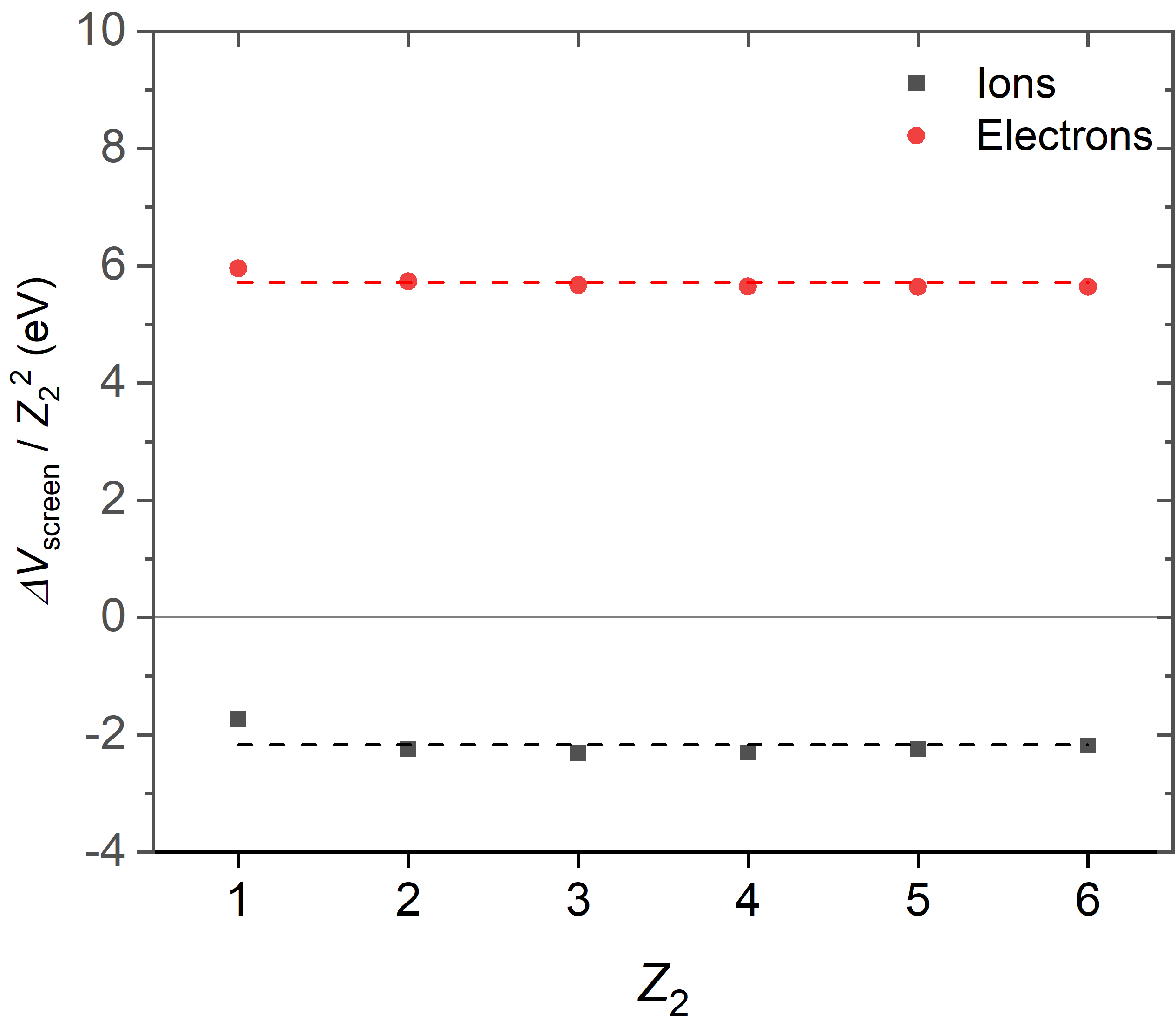}
    \caption{(Color online) Verification of the quadratic projectile charge scaling. Scaled near-field screening potential gap $\Delta V_{\text{screen}} / Z_2^2$ at $r = 20\,\mathrm{fm}$ as a function of projectile charge $Z_2 = 1, 2, \dots, 6$ for a fixed target nucleus $Z_1 = 6$ embedded in a $^{12}$C plasma ($T = 10\,\mathrm{keV}, \rho_{\text{C}} = 240\,\mathrm{g/cm^3}$). Black squares denote the ionic channel, red circles denote the electronic channel, and dashed lines represent constant horizontal fits. The horizontal solid line marks the zero baseline.}
    \label{fig:scaling_z2}
\end{figure}

In contrast, the single-center (SC-TFD) model treats nucleus 1 as the self-consistent polarizing center ($\rho_{\text{ind},1}^{(2)} \propto \chi_2 \Phi_1^2$) and nucleus 2 as a non-polarizing test charge $Z_2 e$. By standard convention for asymmetrical collisions in chapter \ref{sec:theory}, we fix the high-$Z$ target nucleus ($Z_1 \ge Z_2$) at the origin as the self-consistent polarizing center, treating the lighter projectile $Z_2$ as the non-polarizing test charge in the SC-TFD baseline. By electrostatic reciprocity, the single-center interaction represents the pairwise coupling between nucleus 2 and the pre-formed screening cloud of nucleus 1:
\begin{align}
\label{eq:V_sc_order}
V_{\text{screen}}^{\text{SC}(2)}(r) &= -Z_2 e \int G(\mathbf{r}_2, \mathbf{x}) \chi_2 \Phi_1^2(\mathbf{x}) \text{d}^3\mathbf{x} \notag \\
&= -\chi_2 \int \Phi_1^2(\mathbf{x})\Phi_2(\mathbf{x}) \text{d}^3\mathbf{x} \propto - \chi_2 Z_1^2 Z_2.
\end{align}
Equation~\eqref{eq:V_sc_order} proves that the single-center model already captures the $Z_1^2 Z_2$ component of the nonlinear interaction. Subtracting Eq.~\eqref{eq:V_sc_order} from Eq.~\eqref{eq:F_cross}, both the linear Debye terms ($\propto Z_1 Z_2$) and the $Z_1^2 Z_2$ terms cancel identically. The leading-order deviation of the single-center approximation isolates as:
\begin{equation}
\label{eq:master_scaling}
\Delta V_{\text{channel}}(r)  \approx -Z_1 Z_2^2 \cdot  \chi_2^{\text{channel}}(\Gamma, \theta)  \cdot I(r),
\end{equation}
where $I(r) = \int G(\mathbf{x}, \mathbf{r}_1) G^2(\mathbf{x}, \mathbf{r}_2) \text{d}^3\mathbf{x}$ is a spatial overlap integral. 

Equation~\eqref{eq:master_scaling} reveals a factorized scaling law: the two-center correction is governed by an asymmetric quadratic projectile-charge factor $Z_1 Z_2^2$, while the plasma state parameters enter through the effective medium susceptibility $\chi_2^{\text{channel}}(\Gamma, \theta)$. In the asymptotic weak-field limit ($|e\Phi| \ll k_\text{B} T$), the Taylor expansion of the background distributions dictates the onset behavior of $\chi_2$: for classical ions, the Boltzmann expansion yields $\chi_2^{\text{ion}} = \frac{e^3}{2(k_\text{B} T)^2}\sum_i z_i^3 n_{i0} > 0$, scaling with coupling as $\chi_2^{\text{ion}} \propto \Gamma^2$ and imparting a negative sign to $\Delta V_{\text{ion}}$ (screening suppression). For electrons, the Fermi-Dirac expansion yields $\chi_2^{\text{e}} = -\frac{e^3 n_{e0}}{2(k_\text{B} T)^2} \frac{F_{-3/2}(\eta_0)}{F_{1/2}(\eta_0)} < 0$, giving rise to a positive $\Delta V_{\text{e}}$ (screening enhancement).

Figure~\ref{fig:scaling_z2} confirms the quadratic projectile-charge scaling predicted by Eq.~\eqref{eq:master_scaling}. In a benchmark $^{12}$C plasma ($T = 10\,\mathrm{keV}, \rho_\text{C} = 240\,\mathrm{g/cm^3}$), we fix the target nucleus at $Z_1 = 6$ and systematically scan $Z_2 = 1\text{--}6$, extracting the channel-resolved corrections at $r = 20\,\mathrm{fm}$. When scaled by $Z_2^2$, the numerical gaps $\Delta V_\text{screen} / Z_2^2$ collapse into nearly ideal horizontal plateaus across the entire sequence of projectile species, confirming the quadratic dependence $\Delta V_{\text{channel}} \propto Z_2^2$, and the opposite signs of the two plateaus confirm that the electronic accumulation and ionic evacuation act in opposition.

\begin{figure}[t]
    \centering
    \includegraphics[width=0.9\columnwidth]{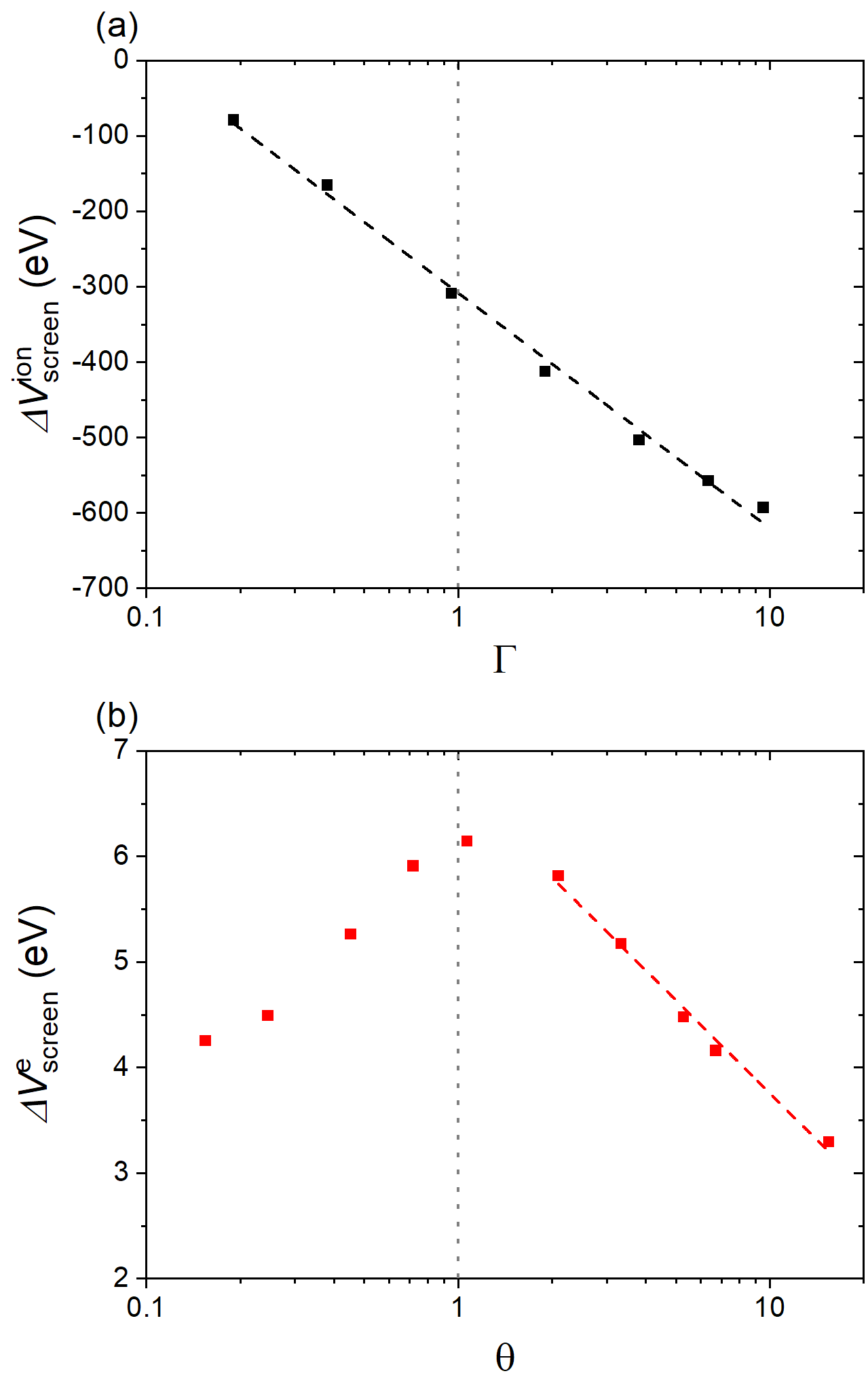}
    \caption{(Color online) Dependence of two-center screening gaps on plasma state parameters, extracted at $r = 20\,\mathrm{fm}$. 
    (a) Ionic screening gap $\Delta V_{\text{screen}}^{\text{ion}}$ as a function of the ion coupling parameter $\Gamma$ for $^{12}$C plasma ($\rho_{\text{C}} = 240\,\mathrm{g/cm^3}$) across temperatures $T = 10\,\mathrm{keV}$ down to $200\,\mathrm{eV}$. The dashed curve shows a linear fit on the semi-log scale, with the vertical dotted line marking the $\Gamma = 1$. 
    (b) Electronic screening gap $\Delta V_{\text{screen}}^{\text{e}}$ as a function of the electron degeneracy parameter $\theta = k_\text{B} T / E_\text{F}$ for a D-T plasma at fixed $T = 1.0\,\mathrm{keV}$ across densities $\rho = 10\,\mathrm{g/cm^3}$ to $5000\,\mathrm{g/cm^3}$. The dashed line represents the classical linear trend fitted to the non-degenerate regime ($\theta \ge 2$), and the vertical dotted line indicates the quantum--classical boundary $\theta = 1$.}
    \label{fig:gamma_theta_scans}
\end{figure}

Having confirmed the nuclear-charge factor, we elucidate the medium thermodynamic response $\chi_2^{\text{channel}}(\Gamma, \theta)$ in Fig.~\ref{fig:gamma_theta_scans}. In Fig.~\ref{fig:gamma_theta_scans}(a), we cool a $^{12}$C plasma at fixed $\rho_\text{C} = 240\,\mathrm{g/cm^3}$ from $10\,\mathrm{keV}$ to $200\,\mathrm{eV}$ (thereby tuning $\Gamma$ from $0.19$ to $9.5$) and extract the channel-resolved ionic screening gap $\Delta V_{\text{screen}}^{\text{ion}}$. The ionic gap $\Delta V_{\text{screen}}^{\text{ion}}$ deepens monotonically, following a linear slope on the semi-log plot. In Fig.~\ref{fig:gamma_theta_scans}(b), we compress a D-T plasma at fixed $T = 1.0\,\mathrm{keV}$ over $\rho = 10\text{--}5000\,\mathrm{g/cm^3}$ (scanning $\theta$ from $16$ down to $0.16$) and extract the corresponding electronic screening gap $\Delta V_{\text{screen}}^{\text{e}}$. In the non-degenerate regime ($\theta \ge 1$), $\Delta V_{\text{screen}}^{\text{e}}$ increases steadily with decreasing $\theta$, well described by the linear fit on the semi-log plot driven by the rising ambient electron density. However, upon traversing the quantum-classical transition boundary $\theta \approx 1$ (vertical dotted line), the electronic enhancement reaches a maximum and turns downward. This turning point reflects the competition between two opposing trends: the ambient electron density increases with compression and supplies more electrons available for screening, whereas Fermi-Dirac exclusion rapidly exhausts available low-energy phase space once the electron becomes degenerate ($\theta \lesssim 1$), suppressing further nonlinear electron pileup into the overlapping potential well. 

In summary, these systematic investigations establish and validate the factorized architecture of the two-center correction.
While the perturbative framework successfully reveals the exact $ Z_2^2$ projectile-charge factor and dictates the opposing signs of the two polarization channels, the thermodynamic response $\chi_2^{\text{eff}}$ transitions from the idealized asymptotic power law ($\propto \Gamma^2$) to logarithmic non-perturbative regimes ($\propto \ln \Gamma$ and $\ln \theta$). 

From a practical perspective, this factorized scaling law provides an efficient framework for dense plasma applications. Fully solving the 2D self-consistent TC-TFD equations across multi-scale grids is computationally demanding. However, our findings demonstrate that the two-center screening potential can be reconstructed by:
\begin{equation}
\label{eq:extrapolation_scheme}
V_{\text{screen}}^{\text{TC}}(r) \approx V_{\text{screen}}^{\text{SC}}(r) + \Delta V_{\text{ion}}(r; Z_1, Z_2, \Gamma) + \Delta V_{\text{e}}(r; Z_1, Z_2, \theta),
\end{equation}
where the single-center baseline $V_{\text{screen}}^{\text{SC}}(r)$ is obtained at negligible computational cost via a standard 1D radial solver. Because $\Delta V_{\text{channel}}$ follows the quadratic $Z_2^2$ trajectory and exhibits smooth, well-defined logarithmic scaling over typical operational ranges of $\Gamma$ and $\theta$, one can readily construct semi-empirical parameterizations for the channel deviations. This enables an instantaneous extrapolation from 1D single-center solutions to accurate 2D two-center screening potentials across diverse thermonuclear plasma compositions.

\begin{figure*}[t]
    \centering
    \includegraphics[width=1\textwidth]{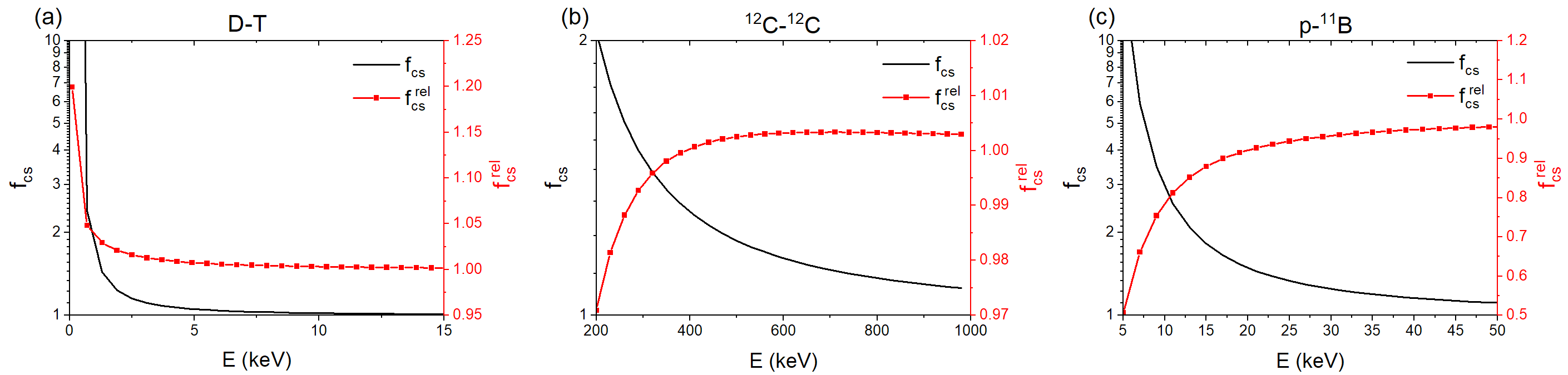}        
    \caption{Enhancement of the fusion cross-section under different plasma conditions. The solid black line shows the absolute enhancement of the TC-TFD model relative to the bare Coulomb potential. The red squares show the relative ratio of the TC-TFD enhancement to the DH prediction. The specific parameters are: (a) D-T mixture at $T = 1\,\mathrm{keV}$ with $\rho_{\text{D}} = 20\,\mathrm{g/cm^3}$ and $\rho_{\text{T}} = 30\,\mathrm{g/cm^3}$; (b) $^{12}$C-$^{12}$C mixture at $T = 10\,\mathrm{keV}$ with $\rho_{\text{C}} = 240\,\mathrm{g/cm^3}$, (c) p-$^{11}$B mixture at $T = 1\,\mathrm{keV}$ with $\rho_{\text{p}} = 20\,\mathrm{g/cm^3}$ and $\rho_{\text{B}} = 220\,\mathrm{g/cm^3}$.
    }
    \label{fig.6}
\end{figure*}

\subsection{The enhancement of fusion cross-sections and reaction rates}

We now apply the self-consistent TC-TFD screening potential to evaluate the cross-section enhancements and thermonuclear reaction rates for the D-T, $^{12}$C-$^{12}$C, and p-$^{11}$B reactions under the plasma conditions listed in Table~\ref{tab:params}. We define the cross-section enhancement factor as $f_\text{cs}(E) \equiv \sigma(E)/\sigma_\text{bare}(E)$, and the relative ratio $f_\text{cs}^\text{rel}(E) \equiv f_\text{cs}^\text{TC-TFD}(E) / f_\text{cs}^\text{DH}(E)$ as a direct measure of the overall nonlinear two-center effects relative to the linear DH benchmark.

Across all three reactions, the black curves in Fig.~\ref{fig.6} show $f_\text{cs}(E)$ decreasing monotonically with incident center-of-mass (CM) kinetic energy $E$ and tending asymptotically to unity. This common behavior is governed by the classical turning point $r_\text{c}$, defined by $E = V_\text{bare}(r_\text{c}) = Z_1 Z_2 e^2 / (4\pi\epsilon_0 r_\text{c})$ and the ratio of $\frac{V_\text{screen}(r_\text{c})}{V_\text{bare}(r_\text{c})} $. As $E$ rises, the screening energy $V_\text{screen}(r_\text{c})$ becomes a negligible fraction of the incident energy, driving $f_\text{cs}$ asymptotically to unity.

The relative ratio $f_\text{cs}^\text{rel}(E)$ (red curves with squares) splits into three distinct behaviors, reflecting which polarization channel dominates at the turning point $r_\text{c}(E)$. For the weakly coupled D-T reaction [Fig.~\ref{fig.6}(a)], $f_\text{cs}^\text{rel}$ remains strictly above unity throughout the Gamow domain, providing an enhancement of nearly $4\%$ over the DH prediction at $E \sim 1\,\mathrm{keV}$. In this regime, the attractive two-center well drives strong nonlinear electron accumulation that outcompetes the background ion evacuation, elevating $V_\text{screen}$ above the linear Debye limit. Conversely, for the highly asymmetric p-$^{11}$B system [Fig.~\ref{fig.6}(c)], $f_\text{cs}^\text{rel}$ is heavily suppressed below unity. Here, the proton fails to accumulate sufficient electrons to counterbalance the Coulomb hole formed around the boron core, depressing the screening potential below the DH baseline. The $^{12}$C-$^{12}$C system [Fig.~\ref{fig.6}(b)] exhibits a distinct energy crossover: $f_\text{cs}^\text{rel}$ is suppressed below unity at low energies but flips above unity for $E > 410\,\mathrm{keV}$. This transition directly maps the spatial profile of $V_\text{screen}(r)$ [Fig.~\ref{Different_models}(b)]: as $E$ rises, the turning point $r_\text{c}(E) \propto 1/E$ sweeps inward from the far-field ion-depleted zone into the near-field electron-concentrated core, reversing the screening deviation from suppression to enhancement.

\begin{figure}[t]
    \centering
    \includegraphics[width=0.9\linewidth]{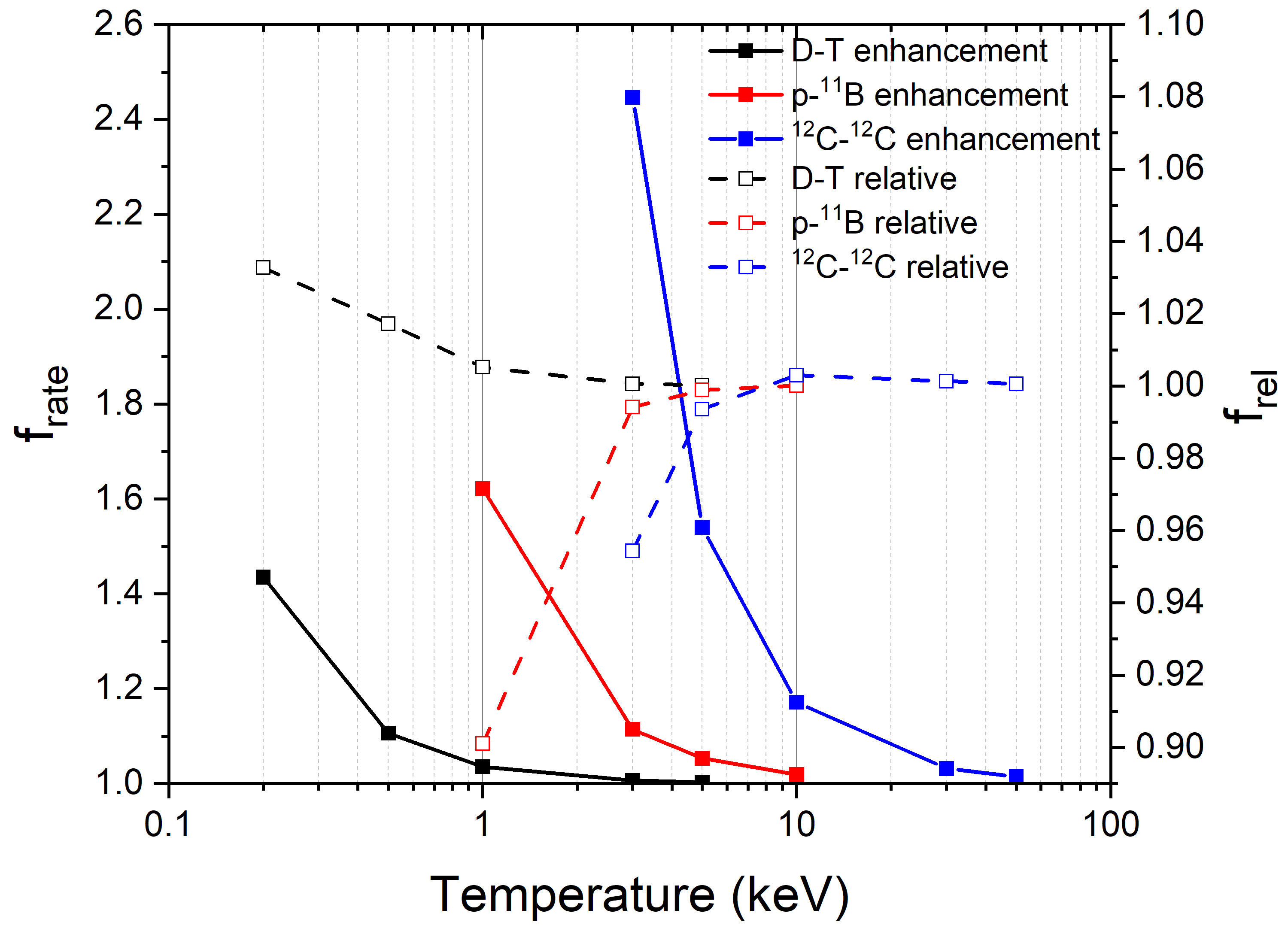}
    \caption{Enhancement of the fusion reaction rate in screening models as a function of temperature. The solid lines (left axis) represent the absolute enhancement factor $f_{\text{rate}}$ relative to the bare nucleus collision. The dashed lines (right axis) indicate the relative ratio of the TC-TFD reaction rate to the classical DH prediction $f_{\text{rel}} = \langle \sigma v \rangle_{\text{TC-TFD}}/\langle \sigma v \rangle_{\text{DH}}$.}
    \label{fig.7}
\end{figure}

The macroscopic consequences of these screening corrections on the thermonuclear reaction rates are presented in Fig.~\ref{fig.7}, showing the rate enhancement factor $f_{\text{rate}} = \langle \sigma v \rangle_{\text{TC-TFD}}/\langle \sigma v \rangle_{\text{bare}}$ and the relative ratio $f_{\text{rel}} = \langle \sigma v \rangle_{\text{TC-TFD}}/\langle \sigma v \rangle_{\text{DH}}$. Because the thermonuclear reaction rate is governed by the Maxwellian-averaged Gamow energy window, increasing the plasma temperature shifts the effective turning point inward, sweeping the barrier sampling from the far field toward the near field. 

For the D-T plasma, the TC-TFD model predicts a reaction rate enhancement consistently higher than the DH prediction ($f_\text{rel} > 1$), exhibiting a nearly $3.3\%$ relative increase at $T = 200\,\mathrm{eV}$ and $2.1\%$ at $500\,\mathrm{eV}$, before decaying asymptotically toward unity at higher temperatures. In contrast, for the p-$^{11}$B plasma, $f_\text{rel}$ remains strictly bounded below unity in the low-temperature regime ($f_\text{rel} \approx 0.965$ at $T = 1\,\mathrm{keV}$), reflecting the dominant far-field ion-evacuation penalty, and recovers monotonically toward unity ($f_\text{rel} \to 1$ at $T \ge 5\,\mathrm{keV}$) as the inward thermal shift alleviates this suppression. The $^{12}$C-$^{12}$C plasma exhibits a distinct non-monotonic turnover in its relative enhancement ratio. At lower temperatures ($T = 5\,\mathrm{keV}$), the Gamow window samples the ion-depleted far field, leading to screening suppression ($f_\text{rel} \approx 0.992 < 1$). As the temperature rises to $10\,\mathrm{keV}$, the Gamow peak advances into the near-field electron-accumulation well, driving $f_\text{rel}$ above unity to $1.004$. Ultimately, at even higher temperatures ($T \ge 30\,\mathrm{keV}$), thermal kinetic energies overwhelm electrostatic screening altogether, causing the relative enhancement ratios for all reactions to asymptotically approach unity ($f_{\text{rel}} \to 1$).

\section{Conclusion and Outlook}

In this work, we have developed a finite-temperature two-center Thomas-Fermi-Dirac screening framework to investigate nonlinear plasma polarization effects on fusion reactions. By evaluating the potential of mean force, our model captures the spatial deformation and merging of the shared screening cloud at the self-consistent field level and obtains the effective screening potential. We demonstrate that this screening potential is governed by an interplay between background-ion evacuation and electron accumulation. Furthermore, the two-center correction is shown to obey a factorized scaling law, $\Delta V_{\text{channel}} \propto - Z_2^2 \cdot \chi_2^{\text{channel}}(\Gamma, \theta)$. Consequently, the resulting thermonuclear reaction-rate enhancements exhibit distinct deviations from the DH predictions: the TC-TFD model yields a persistent rate boost for D-T, a systematic suppression for p-$^{11}$B, and a unique non-monotonic turnover for $^{12}$C-$^{12}$C across temperatures. These findings highlight the limitations of single-center screening models and provide a framework for estimating thermonuclear reaction rates in dense plasma environments.

Despite these capabilities, the present framework entails two primary theoretical approximations. First, the model is formulated at the self-consistent mean-field level. The ionic subsystem is closed by a classical Boltzmann distribution, and the electrons are described within a finite-temperature local-density approximation retaining only the zero-temperature Dirac exchange term. The framework therefore omits multi-ion spatial correlations and the static structure factors $S(k)$ \cite{ichimaru1987statistical,ichimaru1993nuclear} that become prominent in strongly coupled regimes ($\Gamma \gtrsim 1$), as well as finite-temperature exchange-correlation corrections for the electrons. Second, the model relies on the assumption of static thermodynamic equilibrium to determine the self-consistent field distributions, which neglects dynamic non-equilibrium responses during the nuclear collision.

Future investigations will aim to address these limitations and explore further applications of the model. First, we will extend the multi-center framework to explore how high-$Z$ contaminants modulate nonlinear screening and burn degradation at turbulent mix interfaces \cite{berzak2018increasing, casey2023towards, grayson2025nuclear}. Second, benchmarking the TC-TFD screening potentials against first-principles simulations—such as PIMC \cite{pollock2004dense} and QMD \cite{collins1995quantum}—will provide a route to assess many-body correlations beyond the mean-field level. Finally, incorporating our screening potential into quantum barrier penetration calculations in the presence of ambient plasma radiation fields \cite{wu2024multiphoton, thomson2024laser} will allow us to evaluate the combined impact of nonlinear screening and radiation fields on fusion burn dynamics in ICF experiments.

\begin{acknowledgements}
This work was supported by the National Natural Science Foundation of China under Grant No. 12588301, 12375235, 12375240, 12535015 and 12605397.
\end{acknowledgements}

\appendix

\section{Finite-temperature Thomas-Fermi-Dirac Model}
\label{sec:tfd_model}

As discussed in Sec.~\ref{sec:theory}, the equilibrium density distributions of the background plasma are obtained by minimizing the grand potential of the system. To establish the thermodynamic consistency of the TC-TFD screening model, this appendix provides the explicit mathematical formulation of the grand potential and Helmholtz free energy functionals.

Consider the two colliding nuclei separated by an internuclear distance $r$. The grand potential functional $\Omega$ and the Helmholtz free energy functional $F$ of the finite-temperature TFD model are expressed as:
\begin{align}
&\Omega[n_i,n_e] = F[n_i, n_e] -\sum_{i} \mu_i \int n_i(\mathbf{r}) \, \text{d}^3\mathbf{r} \notag \\
&\qquad - \mu_e \int n_e(\mathbf{r}) \, \text{d}^3\mathbf{r},
\label{eq:grand_potential}\\
&F[n_i,n_e] = \sum_i F_{\text{kin}}[n_i] + F_{\text{kin}}[n_e] + E_{\text{ext}}[\rho] \notag \\
&\qquad  + E_{\text{Coulomb}}[\rho] + E_{\text{x}}[n_e],
\label{eq:helmholtz}
\end{align}

where $\rho(\mathbf{r}) = e\bigl(\sum_i z_i n_i(\mathbf{r}) - n_e(\mathbf{r})\bigr)$ is the net induced charge density of the background plasma. The summation over $i$ runs over all background ion species, and $\mu_i$, $\mu_e$ are the asymptotic chemical potentials ensuring charge neutrality at infinity. The non-interacting kinetic free-energy functionals for the classical ions and the partially degenerate electrons are:
\begin{align}
F_{\text{kin}}[n_i] &= k_\text{B} T \int n_i(\mathbf{r}) \left[ \ln\bigl(n_i(\mathbf{r}) \Lambda_i^3\bigr) - 1 \right] \text{d}^3\mathbf{r},
\label{eq:kin_ions}\\
F_{\text{kin}}[n_e] &= k_\text{B} T \int \left[ n_e(\mathbf{r}) \eta_e(\mathbf{r}) - \frac{2A_e}{3} \mathcal{F}_{3/2}\bigl(\eta_e(\mathbf{r})\bigr) \right] \text{d}^3\mathbf{r},
\label{eq:kin_elec}
\end{align}
where $\Lambda_i = h / \sqrt{2\pi m_i k_\text{B} T}$ is the thermal de Broglie wavelength of the $i$-th ion species with mass $m_i$, $A_e = \sqrt{2}(m_e k_\text{B} T)^{3/2} / (\pi^2 \hbar^3)$ is the electron density-of-states prefactor, and $\mathcal{F}_{3/2}$ is the Fermi-Dirac integral of order $3/2$ \cite{blakemore1982approximations}. Within this local functional framework, the local degeneracy parameter $\eta_e(\mathbf{r})$ is implicitly determined by the local electron density via $n_e(\mathbf{r}) = A_e \mathcal{F}_{1/2}\bigl(\eta_e(\mathbf{r})\bigr)$.

The external interaction energy between the background plasma and the two reacting nuclei is:
\begin{align}
E_{\text{ext}}[\rho] = \int \rho(\mathbf{r}) \Phi_{\text{ext}}(\mathbf{r}) \, \text{d}^3\mathbf{r},
\label{eq:external}
\end{align}
where $\Phi_{\text{ext}}(\mathbf{r}) = \frac{e}{4\pi \epsilon_0} \sum_{j=1,2} \frac{Z_j}{|\mathbf{r}-\mathbf{r}_j|}$ is the bare electrostatic potential produced by the colliding pair. The classical Coulomb interaction energy of the polarized plasma reads:
\begin{align}
E_{\text{Coulomb}}[\rho] = \frac{1}{2} \iint \frac{\rho(\mathbf{r}) \rho(\mathbf{r}')}{4\pi \epsilon_0 |\mathbf{r}-\mathbf{r}'|} \, \text{d}^3\mathbf{r} \, \text{d}^3\mathbf{r}',
\label{eq:coulomb}
\end{align}
and the Dirac exchange energy is evaluated in the zero-temperature local density approximation \cite{dirac1930note}:
\begin{align}
E_{\text{x}}[n_e] = -\frac{3}{4} \left( \frac{3}{\pi} \right)^{1/3} \frac{e^2}{4\pi \epsilon_0} \int n_e^{4/3}(\mathbf{r}) \, \text{d}^3\mathbf{r}.
\label{eq:exchange}
\end{align}
Using the zero-temperature form for the exchange functional is standard and robust in finite-temperature TFD models \cite{cowan1957extension}.

Thermodynamic equilibrium requires the grand potential to be stationary with respect to arbitrary variations of the local particle densities. Taking the functional derivatives yields the Euler-Lagrange conditions:
\begin{align}
\frac{\delta \Omega}{\delta n_e(\mathbf{r})} &= k_\text{B} T \eta_e(\mathbf{r}) - e\Phi(\mathbf{r}) + V_\text{x}(\mathbf{r}) - \mu_e = 0, \label{eq:var_e} \\
\frac{\delta \Omega}{\delta n_i(\mathbf{r})} &= k_\text{B} T \ln\bigl(n_i(\mathbf{r}) \Lambda_i^3\bigr) + z_i e \Phi(\mathbf{r}) - \mu_i = 0, \label{eq:var_i}
\end{align}
where $\Phi(\mathbf{r})$ represents the total self-consistent electrostatic potential, and $V_\text{x}(\mathbf{r}) = \frac{\delta E_\text{x}}{\delta n_e} = -\frac{e^2}{4\pi\epsilon_0} [3 n_e(\mathbf{r}) / \pi]^{1/3}$ is the local Dirac exchange potential. Inverting Eqs.~\eqref{eq:var_e} and \eqref{eq:var_i} yields the Fermi-Dirac electron density [Eq.~\eqref{eq:ne}] and the classical Boltzmann ion distribution [Eq.~\eqref{eq:ni}] presented in the main text.

\section{Numerical Method: Recursive Two-Way Coupled Grid-Nesting Algorithm}
\label{app:numerical_method}

Solving the highly nonlinear self-consistent field equations for two approaching nuclei in a dense plasma presents a multi-scale computational challenge. The relevant spatial dimensions span several orders of magnitude: the macroscopic boundary conditions must be imposed at distances covering several Debye lengths ($\lambda_D \sim 10^{-9}\,\mathrm{m}$), whereas the nonlinear polarization and deformation of the screening cloud occur at the sub-atomic internuclear scale ($r_c \sim 10^{-15}\,\mathrm{m}$). Resolving both extremes using a uniform grid would be prohibitive in memory and cost. To circumvent this, we develop a recursive, two-way coupled grid-nesting algorithm in a 2D cylindrical coordinate system $(s, z)$, exploiting the azimuthal symmetry around the internuclear axis.

\subsection{Spatial Discretization and Singularity Separation}
Directly solving the Poisson equation with point-charge nuclei leads to numerical singularities at the grid origins. To resolve this, we decompose the total electrostatic potential $\Phi(\mathbf{r})$ into the known bare nucleus potential and the unknown induced plasma potential:
\begin{align}
    \Phi(\mathbf{r}) = \Phi_{\text{nuclei}}(\mathbf{r}) + \Phi_{\text{plasma}}(\mathbf{r}),
\end{align}
where $\Phi_{\text{nuclei}}(\mathbf{r})$ is the known potential due to the colliding nuclei and $\Phi_{\text{plasma}}(\mathbf{r})$ is the unknown potential due to the induced charges. To handle the $1/r$ divergence, the bare potentials at the locations of the two colliding nuclei are numerically regularized by evaluating them at a half spatial step. Consequently, the discrete Laplacian operator only needs to act on the induced plasma potential. The nonlinear Poisson equation is thus transformed into a system of algebraic equations:
\begin{align}
    \mathbf{A} \boldsymbol{\Phi}_{\text{plasma}} = \mathbf{b}(\boldsymbol{\Phi}_{\text{plasma}} + \boldsymbol{\Phi}_{\text{nuclei}}), \label{eq:matrix_poisson}
\end{align}
where $\mathbf{A}$ is a sparse, constant block-tridiagonal matrix representing the 2D cylindrical Laplacian operator. To ensure matrix symmetry and numerical stability, this operator is formulated using a variable transformation that eliminates the first-order radial derivative and is scaled to be dimensionless. To accelerate the iterative process, $\mathbf{A}$ is pre-factorized using a direct sparse LU decomposition. 

The source vector $\mathbf{b}$ represents the local induced charge density, evaluated using the finite-temperature TFD theoretical framework established in the main text:
\begin{align}
    b_j = -\frac{e \Delta s \Delta z}{\epsilon_0} \left[ n_i(\Phi_j, T) - n_e(\Phi_j, \mu, V_{\text{x},j}) \right], \label{eq:source_term}
\end{align}
where the spatial factor $\Delta s \Delta z $ arises from scaling the discrete Laplacian matrix $\mathbf{A}$ to be dimensionless, following a variable transformation that eliminates the first-order radial derivative in the cylindrical Laplacian. Here, $n_i$ and $n_e$ follow the Boltzmann distribution and the Fermi-Dirac integral $\mathcal{F}_{1/2}(\eta_e)$, respectively. The local Dirac exchange potential $V_{\text{x},j} \propto n_{e,j}^{1/3}$ is dynamically updated based on the local electron density.

The computational domain is constructed as a hierarchy of nested grids $\Omega_k$ ($k = 1, 2, \dots, K$), where $k=1$ is the outermost macroscopic grid and $K$ is the innermost sub-atomic grid. The spatial step size is recursively refined by a factor of $10$ for each subsequent level (e.g., spatial steps $\Delta h_k = \Delta h_{k-1} / 10$, where $\Delta s_k = \Delta z_k \equiv \Delta h_k$).

\subsection{Two-Way Coupled Iteration and Physical Constraints}
For a given internuclear distance $r$, Eq.~(\ref{eq:matrix_poisson}) is solved iteratively across all grid levels. To ensure global physical consistency and charge conservation, the algorithm employs a two-way communication scheme between the coarse grid ($\Omega_{k-1}$) and the fine grid ($\Omega_k$):

\begin{enumerate}
    \renewcommand{\labelenumi}{(\arabic{enumi})}
    \item \textbf{Prolongation (Downward Passing):} 
    The fine grid requires Dirichlet boundary conditions at its outer edges. These boundary values $\boldsymbol{\Phi}_{\text{boundary}}^{(k)}$ are extracted from the converged potential field of the surrounding coarse grid $\Omega_{k-1}$. To prevent unphysical oscillations and ensure the continuity of the electric field at the grid interface, a high-order shape-preserving spline interpolation (modified Akima algorithm) is utilized.

    \item \textbf{Fine Grid Local Relaxation (Picard Iteration):} 
    Within the fine grid $\Omega_k$, the nonlinear system is solved via Picard iterations. During this local relaxation, the electron and ion density distributions are dynamically updated based on the local effective potential. The iteration continues until the local residual $\|\mathbf{A}\boldsymbol{\Phi}_{\text{plasma}} - \mathbf{b}\| < \epsilon_{\text{tol}}$ is satisfied.

    \item \textbf{Restriction (Upward Passing for Charge Conservation):} 
    To maintain global charge conservation, the precise charge density calculated on the fine grid must be fed back to the coarse grid. This is achieved via a volume-weighted restriction operator. For any control volume $(P,Q)$ on the coarse grid, its updated equivalent particle density $\bar{n}_{P,Q}$ is evaluated by the sum of particles in all fine grid cells $(p,q)$ it encloses:
    \begin{align}
        \bar{n}_{P,Q} = \frac{\sum_{(p,q) \in (P,Q)} n_{p,q} V_{p,q}}{\sum_{(p,q) \in (P,Q)} V_{p,q}},
    \end{align}
    where $V_{p,q} = 2\pi s_p \Delta s \Delta z$ is the cylindrical volume element. The projected charge density replaces the central source terms in Eq.~(\ref{eq:source_term}) for the coarse grid $\Omega_{k-1}$.

    \item \textbf{Global Self-Consistency Loop:} 
    With the updated central source terms, the coarse grid $\Omega_{k-1}$ is re-solved. This generates a new set of boundary conditions for the fine grid. This "coarse $\to$ fine $\to$ coarse" bidirectional loop continues until the relative change in the boundary potentials between successive global iterations falls below $10^{-3}$.
\end{enumerate}

\subsection{Under-Relaxation for Numerical Stiffness}
Due to the exponential (Boltzmann) and Fermi-Dirac nonlinearities in the source term $\mathbf{b}(\boldsymbol{\Phi})$, direct fixed-point iteration leads to divergence, often referred to as numerical stiffness. To stabilize convergence, under-relaxation is used for both the potential fields and the boundary conditions. During each iterative step $m$, the updated potential and charge density vector are linearly mixed with their previous states:
\begin{align}
    \boldsymbol{\Phi}^{(m+1)} = \alpha \boldsymbol{\Phi}_{\text{new}} + (1 - \alpha) \boldsymbol{\Phi}^{(m)},
\end{align}
where the mixing parameter $\alpha$ is dynamically chosen between $0.2$ and $0.6$ depending on the grid depth and the local coupling strength. We have confirmed that the converged solution is independent of the specific $\alpha$ schedule within this range.

\subsection{Extraction of the Screening Potential via  Mean-Force Approach}
The numerical framework is combined with the thermodynamic approach established in the main text. Once global self-consistency is achieved across all nested grids for a specific internuclear distance $r$, the local induced electric field gradient at the position of the test nucleus is evaluated using a central difference scheme:
\begin{align}
    \left. \frac{\partial \Phi_{\text{plasma}}}{\partial z} \right|_r = \frac{\Phi_{\text{plasma}}(0, z_{\text{ion}} + \Delta z) - \Phi_{\text{plasma}}(0, z_{\text{ion}} - \Delta z)}{2\Delta z}.
\end{align}
According to the Hellmann-Feynman theorem, this gradient directly corresponds to the mean thermodynamic force $\mathcal{F}(r)$ exerted on the approaching nuclei by the polarized background plasma. 

By continuously shrinking $r$ by spatial steps $\Delta r$ and dynamically regenerating the nested grids, we record the discrete gradient array along the entire internuclear path. The thermodynamic reversible work (potential of mean force) $W_{\text{rev}}$ is then calculated by numerically integrating these discrete gradients:

\begin{align}
    V_{\text{screen}}(r) = \sum_{r_j = r}^{R_{\text{max}}} \left( Z_2 e \left. \frac{\partial \Phi_{\text{plasma}}}{\partial z} \right|_{r_j} \right) \Delta r_j.
\end{align}

This approach bridges the spatial self-consistent field distribution with the two-center screening potential $V_{\text{screen}}(r)$. For the outermost tail region beyond the largest macroscopic grid ($R_{\text{max}}$), an analytical Debye-Hückel boundary condition is smoothly patched to ensure exact asymptotic behavior at infinity.

\subsection{Channel Decomposition and Grid Stitching}
To isolate the individual contributions of background ions and electrons, we exploit the linearity of the converged discrete Laplacian operator $\mathbf{A}$. Once the self-consistent densities $n_i(\mathbf{r})$ and $n_e(\mathbf{r})$ have converged, the source vector is split into ionic and electronic components:
\begin{align}
b_{j}^{\text{ion}} &= -\frac{e \Delta s \Delta z}{\epsilon_0} \sum_i z_i \bigl[ n_i(\mathbf{r}_j) - n_{i0} \bigr], \\
b_{j}^{e} &= \frac{e \Delta s \Delta z}{\epsilon_0} \bigl[ n_e(\mathbf{r}_j) - n_{e0} \bigr].
\end{align}
The Dirichlet boundary conditions are partitioned according to the respective linear polarizabilities:
\begin{equation}
\boldsymbol{\Phi}_{\text{boundary}}^{\text{ion}} = \frac{\kappa_i^2}{\kappa_\text{tot}^2} \boldsymbol{\Phi}_{\text{boundary}}, \quad \boldsymbol{\Phi}_{\text{boundary}}^{e} = \frac{\kappa_e^2}{\kappa_\text{tot}^2} \boldsymbol{\Phi}_{\text{boundary}},
\end{equation}
where $\kappa_i^2 = \sum_i z_i^2 e^2 n_{i0} / (\epsilon_0 k_\text{B} T)$ and $\kappa_e^2 = e^2 n_{e0} / (\epsilon_0 k_\text{B} T)$ and $\kappa_\text{tot}^2 = \kappa_i^2 + \kappa_e^2$. Solving $\mathbf{A} \boldsymbol{\Phi}_{\text{plasma}}^{\text{ion}} = \mathbf{b}^{\text{ion}}$ and $\mathbf{A} \boldsymbol{\Phi}_{\text{plasma}}^{e} = \mathbf{b}^{e}$ independently via direct backward substitution allows for an exact, non-overlapping channel decomposition: $V_{\text{screen}}(r) = V_{\text{screen}}^{\text{ion}}(r) + V_{\text{screen}}^{e}(r)$. 

Finally, the discrete potential segments obtained across the $K$ nested grid levels are stitched head-to-tail using shape-preserving Makima interpolation. The outermost macroscopic tail ($r \ge R_{\text{max}}$) is calibrated against the single-center TFD asymptotic solution to ensure exact zero-energy referencing at infinity.

\bibliography{ref/refs_short}

\end{document}